\documentclass[
final            
  ]
  {aipproc}

\layoutstyle{6x9}

\begin{document}

\def\agile {\emph{AGILE}}
\def\xmm {\emph{XMM-Newton}}
\def\cha {\emph{Chandra}}
\def\comp {\emph{COMPTEL}}
\def\flux {\mbox{erg cm$^{-2}$ s$^{-1}$}}
\def\lum {\mbox{erg s$^{-1}$}}
\def\nh {$N_{\rm H}$}
\def\psr {PSR~B1509--58}


\title{Observing peculiar $\gamma$-ray pulsars with {\emph{AGILE}}}


\author{M. Pilia}{
address={Dipartimento di Fisica, Universit\`a dell'Insubria, Via
          Valleggio 11, I-22100 Como, Italy},
altaddress={INAF--Oss. Astronomico di Cagliari, 
loc. Poggio dei Pini, strada 54, I-09012 Capoterra, Italy},
altaddress={on behalf of the AGILE Team and AGILE Pulsar Working Group} 
} 
 
\author{A. Pellizzoni}{
  address={INAF--Oss. Astronomico di Cagliari, 
loc. Poggio dei Pini, strada 54, I-09012 Capoterra, Italy},
altaddress={on behalf of the AGILE Team and AGILE Pulsar Working Group}
}

\begin{abstract}
The \emph{AGILE} $\gamma$-ray satellite
 provides large sky exposure levels ($\geq$10$^9$ cm$^2$ s
per year on the Galactic Plane) with sensitivity peaking at $E\sim$100 MeV
where the bulk of pulsar energy output is typically released.
Its $\sim$1 $\mu$s absolute time tagging capability makes it
perfectly suited for the study of $\gamma$-ray pulsars.
\emph{AGILE} collected a
large number of $\gamma$-ray photons from EGRET pulsars ($\geq$40,000
pulsed counts for Vela) in two years of observations
unveiling new interesting features at sub-millisecond level in the
pulsars' high-energy light-curves, $\gamma$-ray emission from pulsar glitches
and Pulsar Wind Nebulae.
\emph{AGILE} detected about 20 nearby and energetic pulsars with good
confidence through timing and/or spatial 
analysis.  Among the newcomers we find  pulsars 
with very high rotational energy losses, such as 
the remarkable PSR\,B1509--58
with a magnetic field in excess of 10$^{13}$ Gauss, and
PSR\,J2229+6114 providing a reliable identification for the previously 
unidentified EGRET source 3EG\,2227+6122.
Moreover, the powerful millisecond pulsar B1821--24, in the globular
cluster M28, is detected during a fraction of the observations.
\end{abstract}

\maketitle


\section{Introduction}
Poor $\gamma$-ray pulsar statistics has been a major difficulty in
assessing the dominant mechanism which channels pulsar rotational
energy into high energy emission as well as understanding the
sites where charged particles are accelerated.
The large field of view of the AGILE Gamma-Ray Imaging
Detector (GRID) \cite{tavani09} allows long uninterrupted observations
and  
simultaneous monitoring of tens of nearby radio pulsars belonging to the
``$\gamma$-ray  
pulsar region'' of the $P$--$\dot{P}$ diagram characterized by
$B>2\times10^{11}$ G and  
spin-down energy $\dot{E}_{\rm{rot}}>1.3\times10^{33}$ erg s$^{-1}$
\cite{pellizzoni04}. 
Here we present the results of three years of pulsar observations with \agile.

\section{\emph{AGILE} Observations and Timing}

Pulsar data were collected since the early phases of the mission.
Timing observations suitable for pulsed signal analysis started
in July 2007. 
\emph{AGILE} pointings consisted of long
exposures 
slightly drifting
with respect to the starting pointing direction.
At the end of October 2009 \agile\
started observing in spinning mode due to reaction wheel 
failure, which is not affecting AGILE/GRID sensitivity and pulsar
observations. 
Data screening, particle background filtering and event direction
and energy reconstruction were performed by the \emph{AGILE} Standard
Analysis Pipeline.  We adopted the \emph{AGILE} event
extraction criteria and timing procedures calibrated and
optimized with the observations of known $\gamma$-ray pulsars as
described in \cite{pellizzoni09a}. 

Given the measured \emph{AGILE}'s time tagging accuracy of $<$200 $\mu$s and the
good radio monitoring (i.e. valid epoch range, adequate number of time
of arrivals (ToAs)) available for the majority of our targets,
the most significant pulsed 
signal detection is typically expected within the errors of the radio
ephemeris  
frequency values. In particular, we performed standard epoch
folding and hen  
adequate radio observations covering the time span of the $\gamma$-ray
observations  
are available (i.e. where the $\it{WAVE}$ terms are included in TEMPO2
ephemeris  
files), we account also for the pulsar timing 
noise in the folding procedure as reported in \cite{pellizzoni09a}  over a
frequency  
 range defined according to 3$\sigma$ errors of radio ephemeris.
Pearson's $\chi^2$ statistics is applied to the 10-bin folded pulse profiles 
resulting from each set of spin parameters, yielding the 
probabilities (weighted for the number of trials performed on the data set) 
of sampling a uniform distribution, assessing the significance of the pulsed signal
(sinusoidal or not). 
Furthermore, we verified our timing results also applying bin-independent
parameter-free statistics as the $Z^2_n$ test
and the $H$-test that are typically more sensitive than
$\chi^2$ tests for the search of sinusoidal pulses.
Firstly, for each target we searched for pulsed signal using the
whole available data span. Later, each observation block was
analysed to check for possible flux and/or pulse profile
variability.
We also performed a maximum likelihood analysis (ALIKE task) on 
the \emph{AGILE} data for the regions containing our targets in order to exploit
the instrument's imaging capabilities to assess 
$\gamma$-ray source parameters.
Here, we will focus predominantly on the timing analysis. 

In order to perform \emph{AGILE} timing calibration through accurate
folding and phasing
 a dedicated radio pulsar
monitoring campaign (that will continue during the whole \emph{AGILE}
mission) was undertaken, using the telescopes of the European Pulsar Timing
Array, as well as 
the telescopes of Parkes in Australia and Mt Pleasant in
Tasmania. 
Geminga is a radio-quiet pulsar whose ephemeris can be obtained from the
regular 
X-ray observations by \xmm.
In order to verify the performances of the timing analysis procedure
described above, a crucial parameter to check is
the difference between pulsar rotation parameters derived from radio,
X-ray and $\gamma$-ray data. 
The implementation of the folding
method described in \cite{pellizzoni09a} allowed for a
perfect match between the best period resulting from $\gamma$-ray data
and the period predicted by the radio ephemeris with
discrepancies $\Delta P_{\rm Crab}\sim3\times10^{-12}$ s, comparable
to the period search resolution $r_{\rm Crab}\sim2\times10^{-12}$
s. 
Ignoring timing noise in the folding process would yield discrepancies
(and light curve smearing) which are expected to grow when considering
longer observing time span. Thus the contribution of timing noise
should be considered both in high-resolution
timing analysis and in searching for new $\gamma$-ray pulsars. 

\begin{figure}[ht!]
\includegraphics[height=5.cm,width=5.cm]{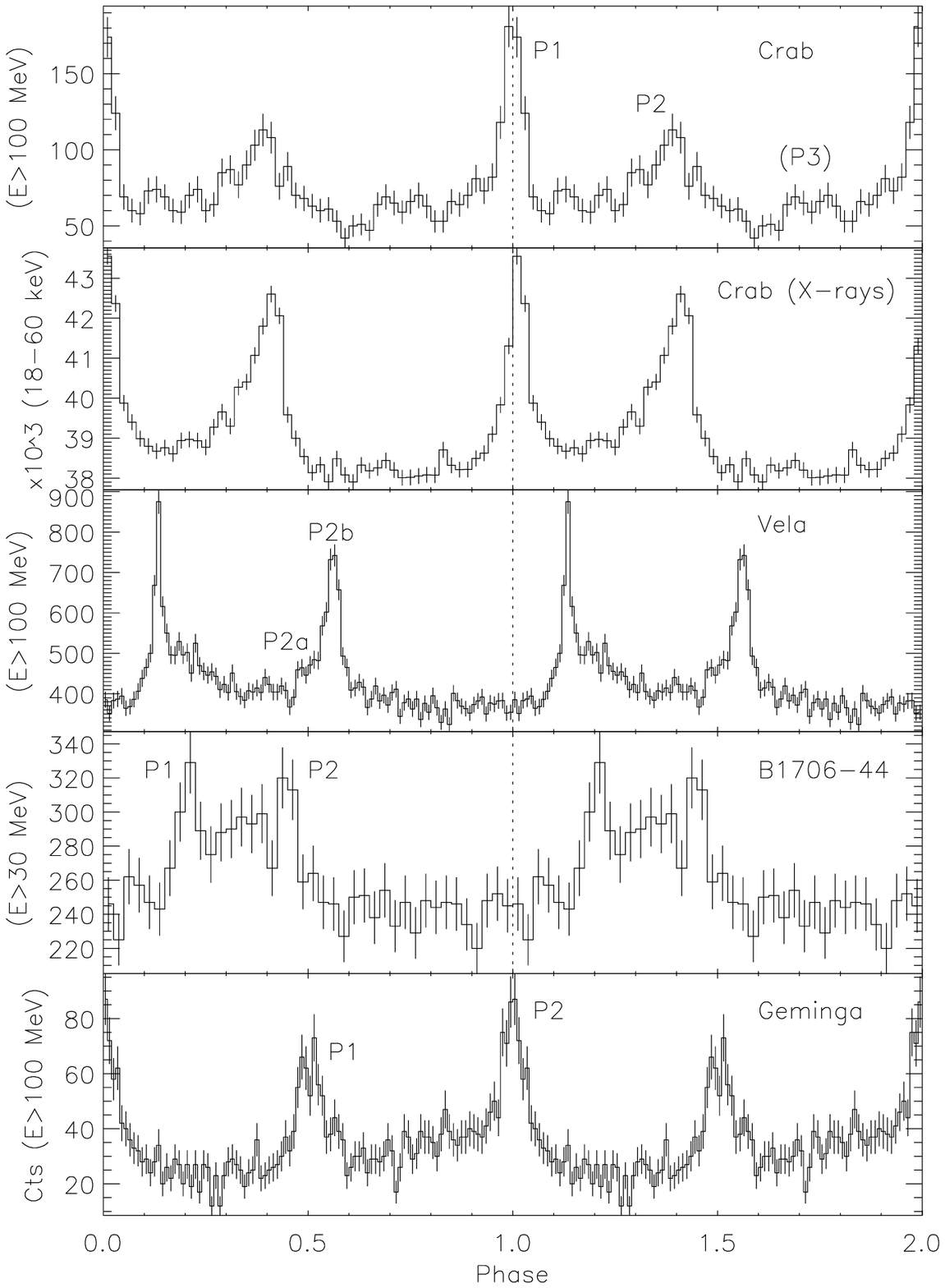}
\includegraphics[height=5.cm,width=5.cm]{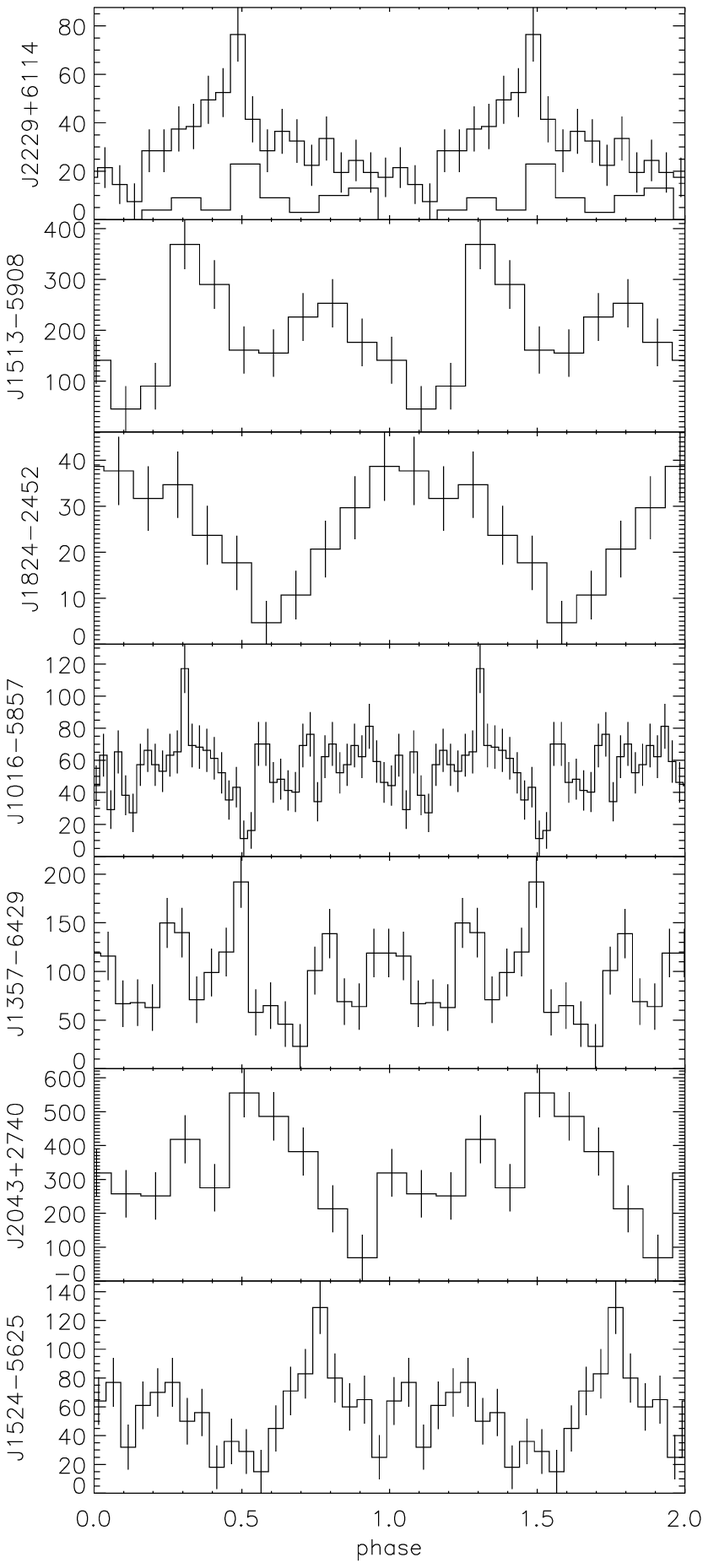}
\caption{\emph{(Left)} Light-curves of the {\emph EGRET} pulsars
obtained
during the first year of \emph{AGILE} scientific operations.
The dashed line corresponds to the the radio main peak.
(see \cite{pellizzoni09a} for details).
\emph{(Right)} Background-subtracted $E>100$ MeV folded pulse profiles 
(except for the $E>50$ MeV J1357--6429 light-curve) of a sample of new
$\gamma$-ray  pulsars discovered by \emph{AGILE}.
For PSR\,J2229+6114, the $E>1$ GeV histogram 
is also shown. 
The main radio peak (1.4 GHz) corresponds to phase 0 (see  \cite{pellizzoni09b} for details).}
\end{figure}

\section{"Old" and New $\gamma$-Ray Pulsars}

The resulting high-energy light-curves
for Vela, Crab, Geminga and PSR~B\,1706--44 pulsars are shown in Figure 1
together with a sample of newly discovered $\gamma$-ray pulsars.

The availability of radio observations bracketing the time span of the
$\gamma$-ray observations (or of X-ray observations very close to the
$\gamma$-ray observations for Geminga) allowed us to also
perform accurate phasing of multi-wavelength light-curves. 
We found that the phasing of the \emph{AGILE} light-curves of the four
pulsars (radio/X-rays/$\gamma$-ray peaks phase separations) is consistent
with EGRET measurements (see \cite{pellizzoni09a}
for details)
implying no evidence of systematic errors in absolute timing with an 
upper limit $t_{err}<1$ ms.
The resulting fluxes (Pulsed
Counts/Exposure) are also consistent with those reported in the EGRET
Catalogue.

The Crab SuperAGILE light-curve (ref) 
was produced with the same folding method reported for the GRID. Inspection of Figure 1 shows that the 
X-ray peaks are aligned with the $E>100$ MeV data within 
$\Delta\phi\sim400$ $\mu$s (a value obtained fitting the peaks with Gaussians) 
providing an additional test of the \emph{AGILE} phasing accuracy.

The effective time resolution of the light-curves will obviously improve with exposure time 
$\Delta t \propto T_{exp}^{-1}$ and a resolution $\leq$50 $\mu$s was 
obtained after two years of \emph{AGILE} observations of Vela.

Although \emph{AGILE} multi-wavelength phasing of the four $\gamma$-ray pulsars is
consistent with the results obtained by \emph{EGRET},
the plots shown in Figure 1 allow us to start assessing new features in 
$\gamma$-ray pulsar light-curves. Narrower and better resolved main peaks 
are revealed, together with previously unknown secondary features.
In particular, a third
peak is possibly detected at $\sim$3.7$\sigma$ level in the Crab light-curve (P3 in Figure 1)
and some interesting features seem present in the Vela light-curves
(confirmed by recent Fermi observations \cite{abdo10_vela}).

In any case, the highly structured light-curves hint at a complex
scenario for the sites of particle acceleration in the pulsar
magnetospheres, implying different electric gaps with physical
properties probably mostly related to their height above the neutron
star surface. Alternatively, slight spatial oscillations of the gap
locations on timescales $\leq$1 day could be invoked to explain the
multiple contiguous peaks seen in the light-curves.

In this perspective, the \emph{AGILE} light-curves time
resolution, currently limited only by the (continuously increasing) source
counts statistics, will eventually yield a pulsar gaps
map (e.g. parametrized with adjustable accelerating electric fields strength,
and location in the magnetosphere) by coupling timing analysis and phase resolved spectral analysis.

The long monitoring of Vela pulsar ($\sim$40,000 pulsed counts in two years
of observations) allowed us to detect possible $\gamma$-ray emission from pulsar
glitches \cite{pellizzoni09a}.

Furthermore, the Vela pulsar wind nebula was recently firmly detected by 
\emph{AGILE} constraining the particle population responsible for the GeV emission and establishing a class of $\gamma$-ray emitters that could account
for a fraction of the unidentified galactic $\gamma$-ray sources \cite{pellizzoni10} (see last Section)

The radio-aligned light-curves of a subset of newly discovered $\gamma$-ray pulsars  for which our timing analysis yielded a 
$>$4$\sigma$ detection are also plotted in Figure 1.
We note that in all cases radio and $\gamma$-ray timing
results are compatible, with the highest significance frequency
detected in $\gamma$-rays within the 
errors of the radio 
ephemeris value, considering also the period search resolution.
Furthermore, we verified that our analysis procedure (potentially affected by instrument-related systematic errors and
biases in events extraction criteria) does not produce fake
detections at a significance level above 3$\sigma$ when the
radio-ephemeris are applied to randomly extracted \emph{AGILE}
data.
The most significant detection is
PSR\,J2229+6114 for which \emph{AGILE} detected pulsed emission
(radio/$\gamma$-ray periods discrepancy $<$$10^{-11}$ s) and pinpointed the most likely position
to $\sim$0.2 deg from the radio pulsar. Our detection provides a
reliable identification for the previously unidentified EGRET source
3EG\, 2227+6122.
The \emph{AGILE} source position,
pulsed flux, and photon
index ($\sim$2.2) are consistent with the EGRET values. The $\gamma$-ray light-curve of this pulsar (detected
up to over 1 GeV),
featuring just one prominent peak shifted $\sim$ 180 deg in phase from the
radio main peak, is shown in Figure 1.

PSR\,J1513--5908 (B1509--58) was detected by COMPTEL in the 1-10 MeV range, while
EGRET reported only marginal evidence for a weak
$<$4$\sigma$ source at $\sim$ 1 deg from the radio position,
with a pulsed emission upper limit
 of $<$$58\times10^{-8}$ ph cm$^{-2}$ s$^{-1}$ (see dedicated section).

At variance with all the other targets, the millisencond pulsar
J1824--2452 in the Globular Cluster M28 was detected by \emph{AGILE},
with good significance ($>$4$\sigma$) and perfect radio-$\gamma$ periods match, 
only in the time interval
54339--54344 MJD.
The main radio peak at 1.4 GHz is coincident with the broad
single peak seen in $\gamma$-rays.
Only marginal
detection was obtained integrating other observations with
comparable exposure or the whole data span. Noise fluctuations
could possibly explain the apparent variability.
Alternatively, although its $\gamma$-ray
efficiency and high-stability of spin parameters are compatible
with rotation-powered emission, some additional mechanism
disturbing the neutron star magnetosphere in the dense cluster
environment could be invoked to explain the variable $\gamma$-ray
phenomenology of this peculiar pulsar. \emph{AGILE} timing failures at
sub-millisecond level in some observations (mimiking source
variability) cannot be excluded. However, this seems unlikely,
since we verified both timing accuracy and stability at $\sim$ 200
$\mu$s level with Vela pulsar observations. Confirmation of this
tantalizing result about physical variability will rest on longer monitoring
campaigns.

\begin{figure}
\includegraphics[width=5cm]{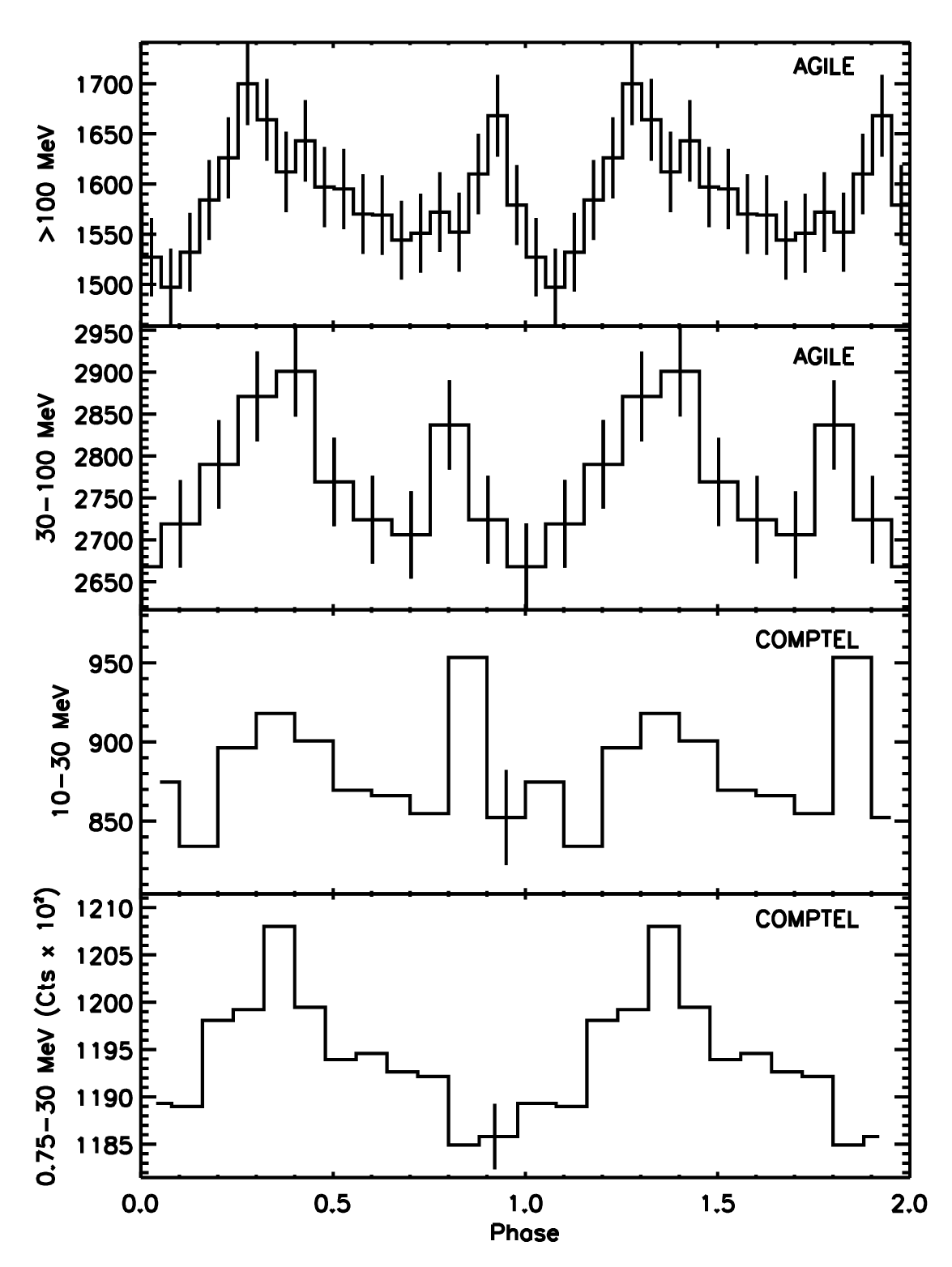} 
\includegraphics[width=6cm]{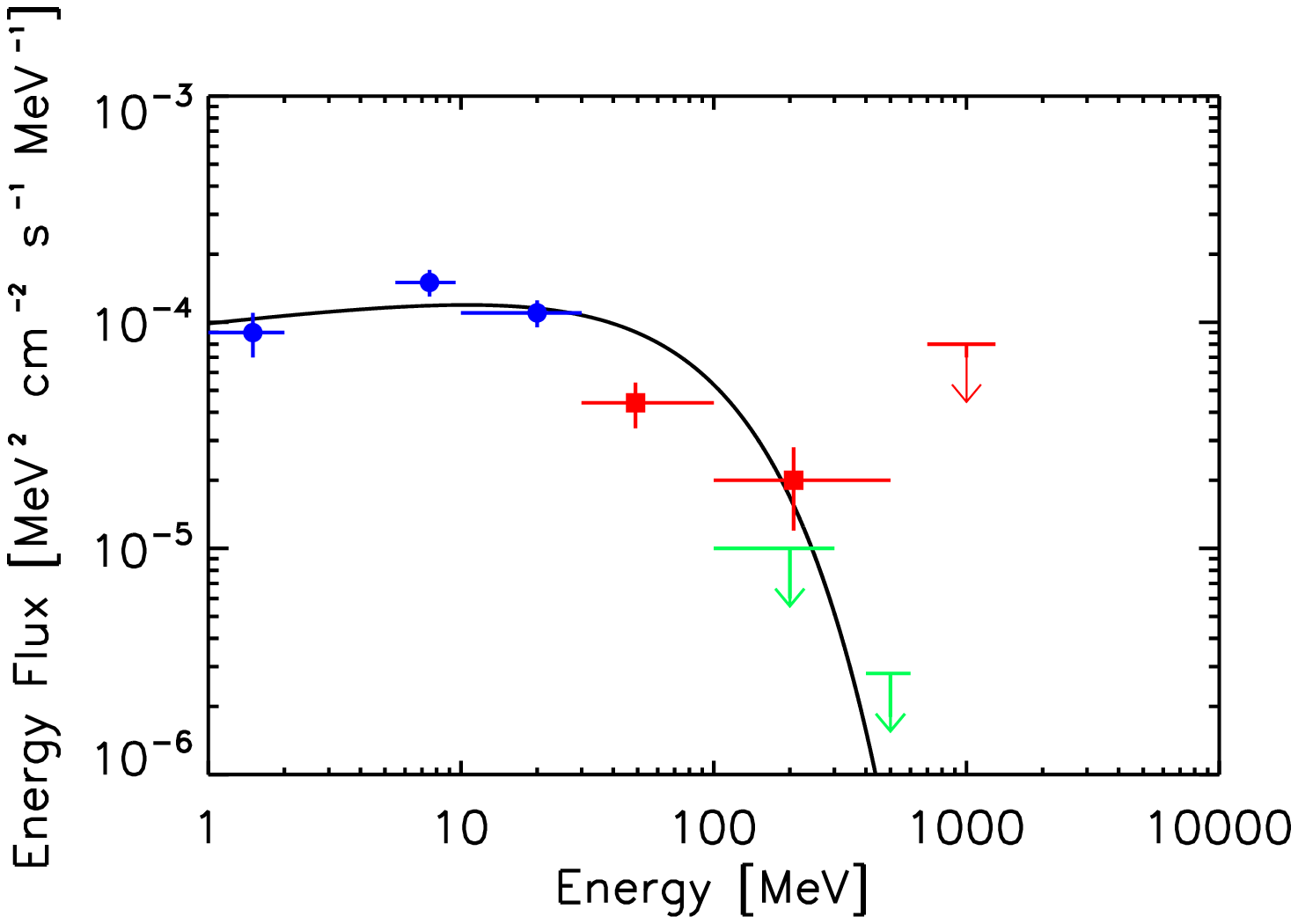}

\caption{\label{fig:lc_tot}
(Left) Phase-aligned $\gamma$-ray light-curves of \psr\ with radio peak at
phase 0. 
(Right) SED of \psr\ (solid
 line) obtained from a fit of pulsed fluxes from soft to hard
  $\gamma$-rays. The circular
   points represent \comp\ observations. The square points
  show  \agile\ pulsed flux at $30<E<100$~MeV and $100<E<500$~MeV. 
The horizontal bar represents
  \agile\ upper limit above 500~MeV. The red arrow represent 
  \agile\ upper limit above 500~MeV.  The two green arrows represent {\it
  Fermi} upper limits \citep{abdo10_1509}. 
\comp\ data are from \cite{kuiper99}. }  
\end{figure}

\section{The ``soft'' PSR~B1509--58: a new class of $\gamma$-ray pulsars?}

\psr\ has a period $P=  150$~ms and a period derivative
$\dot P  =1.53 \times 10^{-12}$s~s$^{-1}$:  assuming the standard dipole
vacuum model, the estimated spin-down 
age for  this pulsar is 1570  years and  its inferred  surface
magnetic field is one of the  highest observed for an ordinary radio pulsar:
$B\approx 
3.1\times 10^{13}$~G, as calculated at the pole; its rotational energy  loss
rate is $\dot E \approx 
1.8 \times  10^{37}$~erg/s. 

The young age  and the  high rotational  energy loss rate made this
pulsar a promising target for the $\gamma$-ray
satellites. In fact, the instruments on board of the
{\it CGRO} observed its
pulsation at low $\gamma$-ray energies,
but it was not detected with high
significance by {\it EGRET}.
\agile\ obtained the first
  detection of 
  \psr\ in the {\it EGRET} band \cite{pellizzoni09b}  confirming the
occurrence of a spectral break. 
We then presented the results of a $\sim 2.5$~yr monitoring campaign
of \psr\ with \agile, improving counts statistics, and therefore
lightcurve characterization, with respect to earlier \agile\ observations
  \cite{pilia10}. 

\agile\ devoted a large amount of observing time to the region of \psr.
We observed \psr\ in three energy bands: 30--100~MeV, 100--500~MeV and above
500~MeV. 
We did not detect pulsed emission at a significance $\sigma \geq 2$ for $E >
500$~MeV. 
The $\gamma$-ray lightcurves of \psr\ for different energy bands
are shown in Fig. \ref{fig:lc_tot}. 
The \agile\ $E>30$ MeV lightcurve shows two peaks. 
The first peak is coincident in phase with \comp's peak \cite{kuiper99}. In its
highest energy band (10--30~MeV) \comp\ showed the indication of a second peak 
(even though the modulation had low significance, $2.1 \sigma$).
This second
peak is coincident in phase with \agile's second peak (Fig. \ref{fig:lc_tot}).
\agile\ thus confirms the previously marginal detection of a second peak.

Based on our exposure 
we derived the $\gamma$-ray flux from the number of
pulsed counts. 
The pulsed fluxes in the three \agile\ energy bands were 
$F_{\gamma}= 10(4)\times 10^{-7}
$~ph~cm$^{-2}$~s$^{-1}$ in the 30--100~MeV band, 
$F_{\gamma}= 2.1(5)\times 10^{-7} $~ph~cm$^{-2}$~s$^{-1}$ in the 100--500~MeV
band  
and a $1 \sigma$ upper limit $F_{\gamma}< 8\times 10^{-8}
$~ph~cm$^{-2}$~s$^{-1}$ for $E>500$~MeV.

Fig. 2 shows the SED of \psr\ based on
\agile's and \comp's observed fluxes. 
\comp\ observations suggested a 
spectral break between 10 and 30 MeV. 
\agile\ pulsed flux at energies $E > 30$~MeV
confirms the presence of a soft spectral break at $E<100$~MeV
As shown in Fig. 2,
we modeled
the observed fluxes with a power-law plus cutoff fit: 
$F(E)=k \times
E^{-\alpha}\exp[-(E/E_{c})^{\beta}]$, 
with three free parameters: the normalization $k$, the spectral index
$\alpha$, the cutoff energy $E_c$ and allowing $\beta$ to assume values of 1
and 2. 
No acceptable $\chi^2$ values were obtained for $\beta=2$,
while for an $\beta=1$ we found $\chi^2_{\nu}=3.2$ for $\nu = 2$
degrees of freedom, corresponding to a
null hypothesis probability of 0.05. 
The best values thus obtained for the parameters of the fit were:
$k=1.0(2)\times 10^{-4}$, $\alpha=1.87(9)$, $E_{c}=81(20)$~MeV.

The bulk of the spin-powered pulsar flux is usually emitted in the MeV-GeV
energy band with  
spectral breaks at $\leq 10$~GeV (e.g. \cite{abdo10psrcat}.
\psr\ has the softest spectrum observed among $\gamma$-ray
pulsars, with a sub-GeV cutoff at $E \approx 80$~MeV. 

When \psr\ was detected in soft $\gamma$-rays but not significantly at $E>30$~MeV,
it was proposed that the mechanism 
responsible for this low-energy spectral break might be photon splitting
\cite{harding97}.
The photon splitting \cite{adler70} is an exotic third-order quantum
electro-dynamics 
process expected when the 
magnetic field approaches or exceeds the $critical$ value defined as
$B_{cr}=m^2_e c^3/(e\hbar)=4.413\times 10^{13}$~G. 
In very
high magnetic fields the formation of pair cascades can be altered
by the process of photon splitting: $\gamma \rightarrow \gamma\gamma$. 

In the case of \psr\ a polar cap model with photon splitting would be
able to explain the soft $\gamma$-ray emission and the low energy
spectral cutoff, now quantified by \agile\ observations.
Based on the observed cutoff, which is related to the photons' saturation
escape energy, 
we can derive constraints on the magnetic field strength at emission,
in the framework of photon splitting:

\begin{equation}
\epsilon_{esc}^{sat} \simeq 0.077(B^{\prime}  \sin \theta_{kB,0})^{-6/5} 
\label{eq:emax}
\end{equation}

where $\epsilon_{esc}$ is the photon saturation escape energy,
$B^{\prime}=B/B_{cr}$ and $\theta_{kB,0} $ is the angle between the 
photon momentum and the magnetic field vectors at the surface and is here
assumed to be very small: 
 $\theta_{kB,0} \leq 0.57 ^{\circ} $
\cite{harding97}. 
Using the observed cutoff ($E= 80$~MeV) we find that $B^{\prime}
\geq 0.3$, which 
implies an emission altitude $\leq 1.3 R_{NS}$, which is the height where
also pair production could ensue.
This altitude of emission is in perfect agreement with the polar cap models
Additionally, \psr\ \cite{kuiper99, crawford01} 
shows evidence of an aligned geometry, which could imply polar cap emission. 

The polar cap model 
as an emission mechanism is debated.
From the theoretical point of view, the angular momentum is
not conserved in polar cap emission (see \cite{treves10} in this book).
And a preferential explanation of the observed $\gamma$-ray
lightcurves with high altitude cascades 
comes from the recent
results by {\it Fermi} \cite{abdo10psrcat}.

Alternatively, 
an interpretation of \psr\ emission can be sought
 in the frame of the three dimensional outer gap model
 \cite{zhangcheng00}. According to these estimates 
a magnetic inclination angle $\alpha\approx 60 ^o$ and a viewing
angle $\zeta \approx 75 ^o$ are
required to reproduce the observed lightcurve. 
Finally, using the simulations of Watters et al. \cite{watters09}, 
the observed lightcurve from \agile\ is best reproduced
if $\alpha\approx 35 ^{\circ}$  and   $\zeta \approx 90
^{\circ}$, in the framework of the two pole caustic model.

The values of $\alpha$ and $\zeta$ required by the model in
\cite{zhangcheng00}  
are not in good 
agreement with the corresponding values obtained with radio measurements.
In fact, Crawford et al. \cite{crawford01} observe that $\alpha$ must be $< 60
^{\circ}$  
at the $3 \sigma$ level.
The prediction obtained by the simulations in \cite{watters09} 
better agrees with the radio polarization 
observations.
In fact, Crawford et al. also propose that,
if the restriction is imposed that $\zeta > 70 ^{\circ}$ \cite{melatos97}, 
then $\alpha > 30 ^{\circ}$ at the $3 \sigma$ level.
For these values, however, the Melatos model for the spin down of an oblique
rotator 
predicts a braking index $n>2.86$, slightly inconsistent with the observed 
value ($n=2.839(3)$).
Therefore, at present the geometry privileged by the
state of the art measurements is best compatible with polar cap models. 

\begin{figure}
\centering
\includegraphics[angle=00,height=5.0cm]{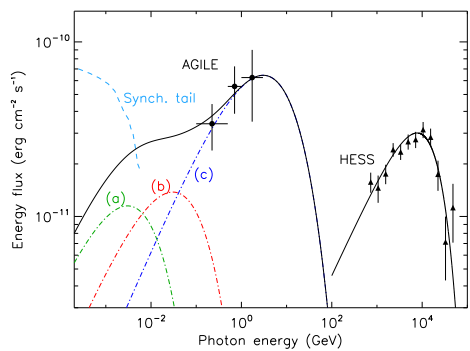}
\caption{Gamma-ray high- and very-high energy spectral  distribution ($\nu
  F_{\nu}$) of the Vela X
PWN. HESS data fit an IC process (scattering on CMBR) related to electron
power law index 2.0 with a break at 67 TeV and a total energy content of
$2.2\times10^{45}$ erg \cite{aharonian06}. 
AGILE data are compatible with IC emission from the additional
electron component
well reproducing the observed total radio spectrum ($E_{\rm tot}=4\times10^{48}$
 erg)
assuming the same $\sim5$ $\mu$G field strength as required
by the TeV spectral break.}
\end{figure}

\section{Pulsars and their environments}

The Vela supernova remnant (SNR) is the nearest SNR ($\sim$290 pc) containing 
a bright pulsar, PSR B0833-45.
The detection of Vela X at TeV energies demonstrated that this
source emits non-thermal radiation, in agreement with the hypothesis that it corresponds
to the pulsar wind nebula, displaced to the south by the unequal pressure of the reverse
shock from the SNR \cite{blondin01}.
The multiwavelength spectrum of the center of Vela X can be modeled as synchrotron radiation
from energetic electrons within the cocoon (radio and x-rays) and
inverse Compton emission from the scattering (by the same
electron population) of the cosmic microwave background
radiation (CMBR), the Galactic far-infrared radiation (FIR) produced by reradiation of dust grains, and the local starlight \cite{lamassa08,dejager08}.
Alternatively, a hadronic model can be invoked for the $\gamma$-ray
emission from the Vela X cocoon, where the emission is the result of the decay of
neutral pions produced in proton-proton collisions \cite{horns06}.
Observations in the MeV-GeV band (HE) are crucial to distinguish between leptonic and hadronic models as well as to identify specific particle populations and spectra. 

With the aim of performing a sensitive search for close faint sources excluding the bright emission from the Vela pulsar, we discarded the time intervals corresponding to the phase interval 0.05--0.65.
The analysis of the resulting off-pulse images unveiled few $\gamma$-ray sources,
none of which coincides with the Vela 
pulsar. A maximum likelihood analysis, performed on the $E>100$ MeV dataset within a region of
5$^{\circ}$ around the pulsar position, revealed two sources at better than
3$\sigma$ confidence, the brightest of which, AGL J0834--4539
($\sim$5.9$\sigma$ significance, $\sim$264 counts, $F_{\gamma}=(35\pm7)\times10^{-8}$ ph cm$^{-2}$ s$^{-1}$ at $E>100$ MeV), is 
positionally concident with HESS J0835--455, the TeV source
that is identified with the Vela X nebula \cite{aharonian06}, and has a similar brightness
profile to it. This implies that AGL J0834--4539 is associated with the
pulsar's PWN.

The AGILE spectral points are a factor $\sim$2 below the previous EGRET upper limits \cite{dejager96} and  well above the extrapolation of
HESS $\nu F_{\nu}$ spectrum to lower energies.
In the frame of leptonic models, the AGILE measurements are not consistent with a simple multiwavelength spectral energy distribution
involving a single electron population.
The AGILE spectral points are one order of magnitude above the fluxes expected from the electron population  simultaneously fitting synchrotron x-ray emission (peaking at $\sim$1 keV)
and inverse-Compton (IC) TeV emission \cite{lamassa08}.
\indent Additional electron populations should be invoked to explain the observed GeV
fluxes. This is not surprising in view of the complex morphology of the PWN
seen in radio and x-rays, where different sites and features of non-thermal
emission are present.
In fact, the position where AGILE sees the maximum brigthness 
is also roughly where the 8.4 GHz radio emission is brightest \cite{hales04}. AGILE data are compatible with the IC parameters modelled by de Jager et al.
 (electron spectral index 1.78
and maximum energy $\sim$20 GeV), although our measurements could suggest a higher contribution from IC photon seeds.
AGILE measurements would be incompatible with the scenario of nucleonic $\gamma$-ray production
in the Vela TeV nebula in the frame of a single primary electron population.

\def\apj {ApJ}
\def\aap {A\&A}
\def\mnras {MNRAS}
\def\apjl {ApJL}
\def\aj {AJ}
\def\apjs {ApJS}
\def\nat {Nature}
\def\aaps {AAPS}
\def\prl {PhRL}

\bibliographystyle{aipproc}   
\bibliography{biblio}

\begin{thebibliography}{23}
\expandafter\ifx\csname natexlab\endcsname\relax\def\natexlab#1{#1}\fi
\providecommand{\enquote}[1]{``#1''}
\expandafter\ifx\csname url\endcsname\relax
  \def\url#1{\texttt{#1}}\fi
\expandafter\ifx\csname urlprefix\endcsname\relax\def\urlprefix{URL }\fi
\providecommand{\eprint}[2][]{\url{#2}}


\bibitem[{Tavani} et~al.(2009)]{tavani09}
M.~{Tavani}, {et al.}
, \emph{\aap} \textbf{502}, 995--1013 (2009).

\bibitem[{Pellizzoni} et~al.(2004)]{pellizzoni04}
A.~{Pellizzoni},   {et al.}, \emph{\apjl} \textbf{612},
  L49--L52 (2004).

\bibitem[{Pellizzoni} et~al.(2009{\natexlab{a}})]{pellizzoni09a}
A.~{Pellizzoni}, 
  {et al.}, \emph{\apj} \textbf{691}, 1618--1633 (2009{\natexlab{a}}),
  

\bibitem[{Pellizzoni} et~al.(2009{\natexlab{b}})]{pellizzoni09b}
A.~{Pellizzoni}, 
  {et al.}, \emph{\apjl} \textbf{695}, L115--L119
  (2009{\natexlab{b}}).

\bibitem[{Abdo} et~al.(2010{\natexlab{a}})]{abdo10_vela}
A.~A. {Abdo}, 
{et al.}, \emph{\apj} \textbf{713}, 154--165
  (2010{\natexlab{a}}).

\bibitem[{Pellizzoni} et~al.(2010)]{pellizzoni10}
A.~{Pellizzoni}, 
{et al.},
  \emph{Science} \textbf{327}, 663-- (2010).

\bibitem[{Abdo} et~al.(2010{\natexlab{b}})]{abdo10_1509}
A.~A. {Abdo}, {et al.}, \emph{\apj} \textbf{714}, 927--936
  (2010{\natexlab{b}}).

\bibitem[{Kuiper} et~al.(1999)]{kuiper99}
L.~{Kuiper}, {et al.}
    , \emph{\aap}
  \textbf{351}, 119--132 (1999).
  

\bibitem[{Pilia} et~al.(2010)]{pilia10}
M.~{Pilia}, {et al.}
, \emph{\apj}
  \textbf{723}, 707--712 (2010).

\bibitem[{Abdo} et~al.(2010{\natexlab{c}})]{abdo10psrcat}
A.~A. {Abdo}, {et al.}, \emph{\apjs} \textbf{187}, 460--494 (2010{\natexlab{c}}).

\bibitem[{Adler} et~al.(1970)]{adler70}
S.~L. {Adler}, J.~N. {Bahcall}, C.~G. {Callan}, and M.~N. {Rosenbluth},
  \emph{PhRL} \textbf{25}, 1061--1065 (1970).

\bibitem[{Harding} et~al.(1997)]{harding97}
A.~K. {Harding}, M.~G. {Baring}, and P.~L. {Gonthier}, \emph{\apj}
  \textbf{476}, 246--+ (1997).

\bibitem[{Crawford} et~al.(2001)]{crawford01}
F.~{Crawford}, R.~N. {Manchester}, and V.~M. {Kaspi}, \emph{\aj} \textbf{122},
  2001--2007 (2001).

\bibitem[{Treves} et~al.(2010)]{treves10}
A.~{Treves}, M.~{Pilia}, and M.~{Lopez Moya}, \emph{ArXiv e-prints}  (2010),
  \eprint{1011.6562}.

\bibitem[{Zhang} and {Cheng}(2000)]{zhangcheng00}
L.~{Zhang}, and K.~S. {Cheng}, \emph{\aap} \textbf{363}, 575--584 (2000).

\bibitem[{Watters} et~al.(2009)]{watters09}
K.~P. {Watters}, R.~W. {Romani}, P.~{Weltevrede}, and S.~{Johnston},
  \emph{\apj} \textbf{695}, 1289--1301 (2009).

\bibitem[{Melatos} (1997)]{melatos97}
A. Melatos, \emph{\mnras} \textbf{288}, 1049+ (1997).

\bibitem[{Blondin} et~al.(2001)]{blondin01}
J.~M. {Blondin}, R.~A. {Chevalier}, and D.~M. {Frierson}, \emph{\apj}
  \textbf{563}, 806--815 (2001).

\bibitem[{LaMassa} et~al.(2008)]{lamassa08}
S.~M. {LaMassa}, P.~O. {Slane}, and O.~C. {de Jager}, \emph{\apjl}
  \textbf{689}, L121--L124 (2008).

\bibitem[{de Jager} et~al.(2008)]{dejager08}
O.~C. {de Jager}, P.~O. {Slane}, and S.~{LaMassa}, \emph{\apjl} \textbf{689},
  L125--L128 (2008).

\bibitem[{Horns} et~al.(2006)]{horns06}
D.~{Horns}, 
{et al.}, \emph{\aap} \textbf{451}, L51--L54 (2006).

\bibitem[{de Jager} et~al.(1996)]{dejager96}
O.~C. {de Jager}, A.~K. {Harding}, P.~{Sreekumar}, and M.~{Strickman},
  \emph{\aaps} \textbf{120}, C441+ (1996).

\bibitem[{Hales} et~al.(2004)]{hales04}
A.~S. {Hales}, {et al.}, \emph{\apj}
  \textbf{613}, 977--985 (2004).

\bibitem[{Aharonian} et al. (2006)]{aharonian06}
F. {Aharonian}, {et al.}, \emph{\aap}
\textbf{448}, L43+, (2006)

\end{thebibliography}

\end{document}